\documentclass[11pt]{article}
\input{epsf}
\newcommand{\beq}{\begin{equation}}
\newcommand{\eeq}{\end{equation}}
\newcommand{\beqa}{\begin{eqnarray}}
\newcommand{\eeqa}{\end{eqnarray}}

\newcommand{\vs}{\vspace{-0.25cm}}

\newcommand{\tc}{{\mathcal{T}}}
\begin{document}

{\tiny \hfill FZJ-IKP(TH)-2000-26}

\vspace{2cm}

\begin{center}

{{\Large \bf 
Chiral dynamics in the presence of bound states: \\[0.3em] 
kaon--nucleon interactions revisited}}

\end{center}

\vspace{.3in}

\begin{center}
{\large
J.A. Oller\footnote{email: j.a.oller@fz-juelich.de}, 
Ulf-G. Mei{\ss}ner\footnote{email: Ulf-G.Meissner@fz-juelich.de}}

\bigskip

{\it Forschungszentrum J\"ulich, Institut f\"ur Kernphysik
(Theorie)\\ D-52425 J\"ulich, Germany}

\bigskip

\end{center}

\vspace{.7in}

\thispagestyle{empty}

\begin{abstract}
\noindent
We study the S--wave kaon--nucleon interactions for strangeness $S=-1$ in a 
novel relativistic chiral unitary approach based on coupled channels. 
Dispersion relations are used to perform the necessary resummation of 
the lowest order relativistic chiral Lagrangian. A good description of the 
data  in the $K^- p$, $\pi \Sigma$ and $\pi \Lambda$ channels
is obtained. We show how this method can be systematically extended 
to higher orders, emphasizing its applicability to any 
scenario of strong self--interactions where the perturbative 
series diverges even at low energies. Discussions 
about the differences to existing approaches employing pseudo--potentials in a regulated 
Lippmann--Schwinger equation are included. Finally, we describe the resonance 
content of our meson--baryon amplitudes and discuss its nature.
\end{abstract}

\vspace{1in}

\vfill

\pagebreak


\medskip

\noindent {\bf 1.} 
The $\bar{K}N$ ($K^- p, \bar{K}^0 n$) interaction is of interest for nuclear, particle and
astrophysics. It is characterized by large rescattering 
effects between different channels and by the 
presence of the $\Lambda(1405)$ resonance just below the  $\bar{K}N$ threshold. 
{}From the theoretical point of view one  expects that chiral symmetry severely 
constrains the interactions between the different channels. However, 
because of such large rescattering effects, for short unitarity corrections, 
pure meson--baryon chiral perturbation theory (CHPT)
\cite{gl,hb1,hb2,hb3,et,bech}  cannot 
be applied. In particular, the 
resonance $\Lambda(1405)$ can be never reproduced in such a perturbative 
framework at any finite order because it is not perturbative in the chiral 
counting\footnote{Of course, one can couple in an explicit $\Lambda(1405)$
field, see e.g. \cite{korea,savage}, but in that case a consistent power
counting does not exist.}. As a consequence of that, a proper way of resummation of 
these strong unitarity corrections in the chiral expansion is necessary. A similar 
situation can be found in the nucleon--nucleon system where the presence of 
shallow nuclear bound states invalidates the application of perturbative effective 
field theory techniques. The solution advocated in ref.\cite{wein} is to apply 
the chiral expansion to generate an ``effective potential'' which is then iterated in a 
Lippmann-Schwinger equation to calculate the whole S--matrix.
Such ideas have been successfully pursued for $\bar{K}N$ scattering based on chiral 
Lagrangians and coupled channel pseudo--potentials in regulated Lippmann--Schwinger 
equations, see \cite{kaiser,oset}. In these studies the nonperturbative nature of the 
strangeness $S= -1$ S--wave kaon--nucleon interaction was clearly established. However, 
from the theoretical point of view, the methods employed in
\cite{kaiser,oset} should further
be improved. First, although chiral perturbation theory Lagrangians are used, 
a direct matching to the proper CHPT {\em amplitudes} is not done, an ingredient 
already stressed a decade ago in a different context \cite{gm}. Second, the results
presented in \cite{kaiser,oset} show a strong sensitivity to the cut--off or regulator
masses. This can be avoided e.g. by employing subtracted dispersion relations,
with the added advantage that the subtraction constants can even be taken at
some unphysical points, where they can be constrained by chiral symmetry (a nice
example for this is the study of higher order corrections in eta decays in 
\cite{kww,la}). Note that the usefulness of considering a subtraction
scheme rather than a cut--off has also been stressed in \cite{lutz}. 
By construction, our approach is fully relativistic and thus one never 
has to make recourse to any expansion in the inverse of the baryon mass, although that 
can be done if desired.  Third, the explicit inclusion of resonance
fields in these schemes is not at all obvious. This becomes of importance if one
wants to decide whether a resonance is simply generated by the strong meson--baryon
dynamics or also has a ``preexisting'' (quark model) component.
Such questions have e.g. been addressed in the context of potential
models of varying sophistication, see \cite{siegel,juelich}. 
We present here a 
general and alternative scheme to that of introducing a potential 
\cite{wein} in order to still apply  chiral Lagrangians to those situations where 
the perturbative chiral expansion fails because of the strong self--interactions 
between the relevant degrees of freedom. Note that these strong interactions 
can even generate poles (e.g. bound states) in the S-matrix as it happens e.g. in some 
channels of the $\bar{K}N$, nucleon--nucleon and even of the meson--meson systems. 
The approach is based on 
coupled channel subtracted dispersion relations and on matching the
general expression for the pertinent partial wave amplitudes
to the results of any given CHPT calculation in a well defined 
chiral power counting.
The method is suited to include the contributions of explicit resonance 
fields, if desired. It is a natural extension and reformulation of 
our previous work on pion--nucleon scattering \cite{piN}. 
Although our approach is more general, in this letter we only show that
even starting from the lowest order tree--level amplitudes of the kaon--nucleon system,
one can fairly well describe the threshold branching ratios and scattering data as well 
as the event distributions (note that we have developed an improved method for 
calculating these event distributions as shown below). We also give a detailed 
comparison to the existing approaches to underline the remarks made in this 
introduction.

\medskip \noindent {\bf 2.} The lowest order ${\cal O}(p)$ SU(3) meson--baryon Lagrangian 
compatible with chiral symmetry and its breakings, both spontaneous and explicit, and 
with parity and charge conjugation can be written as:\footnote{We denote by $p$ any
small parameter like a meson mass or external momentum.}
\beq
\label{lag}
{\mathcal{L}}=\langle i\bar{B}\gamma^\mu \partial_\mu B-m_0 \bar{B}B+\frac{1}{2}D\,
\bar{B}\gamma^\mu \gamma_5\left\{u_\mu,B \right\}+\frac{1}{2}F\,\bar{B}\gamma^\mu 
\gamma_5 [u_\mu,B] \rangle
\eeq
where $m_0$ stands for the (average) octet baryon mass in the chiral limit. The trace
$\langle \ldots \rangle$  runs over the flavor indices and the 
axial-vector couplings $F$ and $D$ are subject to the 
constraint $F+D=g_A= 1.26$. We use the values $D = 0.8$ and $F =
0.46$ extracted from hyperon decays \cite{rat}. Furthermore, we have $u_\mu=i u^\dagger 
(\partial_\mu U) u^\dagger=u_\mu^\dagger$, 
$U(\Phi)=u(\Phi)^2=\exp(i \sqrt{2}\Phi/F_0)$, with $F_0$ the pseudoscalar decay constant 
in the chiral limit. The $3\times 3$ flavor--matrices $\Phi$ and $B$ are given by:
\beqa
\Phi=\sum_i \frac{\lambda_i}{\sqrt{2}}\phi_i &=&\left(
\begin{array}{ccc}
\frac{1}{\sqrt{2}}\pi^0 +\frac{1}{\sqrt{6}}\eta & \pi^+ & K^{+} \\
 \pi^-&-\frac{1}{\sqrt{2}}\pi^0+\frac{1}{\sqrt{6}}\eta& K^{0}\\
 K^{-}& \bar{K}^{0}&-\frac{2}{\sqrt{6}}\eta
  \end{array}
\right) ~,\nonumber \\
B=\sum_i \frac{\lambda_i}{\sqrt{2}}B_i &=&\left(
\begin{array}{ccc}
\frac{1}{\sqrt{2}} \Sigma^0 +\frac{1}{\sqrt{6}}\Lambda & \Sigma^+ & p \\
 \Sigma^-&-\frac{1}{\sqrt{2}}\Sigma^0+\frac{1}{\sqrt{6}}\Lambda& n\\
 \Xi^{-}& \Xi^{0}&-\frac{2}{\sqrt{6}} \Lambda
  \end{array} \right) ~,
\eeqa
where the $\lambda_i$ are the usual SU(3) Gell-Mann matrices.  Notice that for our 
purpose, it is not necessary  to consider in eq.(\ref{lag}) external fields 
related i.e. to the electroweak interactions. 
{}From eq.(\ref{lag}) it is straightforward to write down the lowest order 
CHPT (i.e. tree level) $\phi_i B_a \rightarrow \phi_j B_b$ amplitude,
denoted $T(ij,ab)$. It  is given  by the sum of the diagrams in fig.1:
\beqa
\label{amp}
T(ij,ab)&=&T_s(ij,ab)+T_d(ij,ab)+T_c(ij,ab) ~,\\
T_s(ij,ab)&=&\sum_{k=1}^8 f_{abk}f_{ijk} \frac{N N'}{F_0^2} \chi_b 
\left[W-m_0+\left( \vec{p}\cdot \vec{p}\,'+i(\vec{p}\,'\times \vec{p})\cdot 
\vec{\sigma}\right)\frac{W+m_0}{(N N')^2} \right]\chi_a ~, \nonumber \\
T_d(ij,ab)&=&-\sum_{k=1}^8 (Dd_{jkb}+i Ff_{jkb})(Dd_{ika}-i F f_{ika})
\frac{N N'}{F_0^2}\chi_b \left[  -2m_0+\frac{s+3m_0^2}{W+m_0}
 \right .\nonumber \\
&+&\left . \frac{\vec{p}\cdot\vec{p}\,'+i(\vec{p}\,'\times \vec{p})\cdot \vec{\sigma}}
{(N N')^2}
\left(2m_0+\frac{s+3m_0^2}{W-m_0} \right)\right]\chi_a\nonumber ~, \\
T_c(ij,ab)&=&-\sum_{k=1}^8(D d_{ikb}+i F f_{ikb})(D d_{jka}-i F f_{jka})
\frac{NN'}{F_0^2}\chi_b \left[-2m_0-\frac{(W-m_0)(u+3m_0^2)}{u-m_0^2}
\right . \nonumber \\ &+& \left .
\frac{\vec{p}\cdot\vec{p}\,'+i(\vec{p}\,'\times \vec{p})\cdot \
\vec{\sigma}}{(NN')^2} \left(2 m_0-\frac{W+m_0}{u-m_0^2}(u+3 m_0^2) 
\right) \right]\chi_a ~, \nonumber 
\eeqa
\begin{figure}[t]
\centerline{
\epsfxsize=11cm
\epsffile{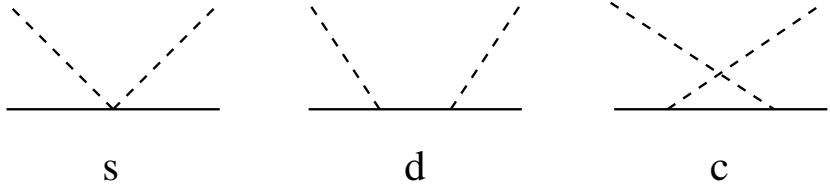}
}
\caption{Leading order tree level diagrams. Baryons and meson are depicted by 
solid and dashed lines, respectively. Shown are the seagull (s),
direct (d) and crossed (c) graphs, in order.}
\vspace{-0.3cm}
\end{figure}

\noindent
where $f_{klm}$ and $d_{klm}$ are constants defined in terms of the commutators and 
anticommutators of the SU(3) Gell-Mann matrices in the usual way: 
$[\lambda_k,\lambda_l] = 2 i \sum_{m=1}^8f_{klm}\lambda_m$,
$\{\lambda_k,\lambda_l\} = 2 \sum_{m=1}^8 d_{klm} \lambda_m$, $W\equiv \sqrt{s}$ is the 
total centre--of--mass (c.m.) energy, and $u=(p-q')^2$, with $p$ and $q'$ the 
four--momentum of the initial baryon and final meson, respectively. In addition, 
$N=\sqrt{E_0+m_0}$ and $N'=\sqrt{E'_0+m_0}$, with $E_0$ and $E'_0$ 
the c.m. energies of the initial and final baryons with masses 
$m_0$, in order. 
Finally, $\vec{p}$ and $\vec{p}\,'$ 
are the c.m. three-momentum vectors of the baryons in the initial and final state, 
respectively and $\chi_c$, with $c=a$ or $b$, is a two component Pauli spinor. To 
calculate the previous expressions, we have considered one particle states 
$|\vec{p},\sigma\rangle$ with c.m. three--momentum $\vec{p}$ and spin component  
$\sigma$ normalized  such that
$\langle \vec{p} ,\sigma |\vec{p}\,',\sigma ' \rangle=2 p^0 (2 \pi)^3
\delta^{(3)}(\vec{p}-\vec{p}\,') \delta_{\sigma \sigma'}$. 
Here, $p^0$ is the energy of the particle,  $\delta^{(3)}(\vec{p}-\vec{p}\,')$ 
the Dirac delta function and $\delta_{\sigma \sigma'}$ the usual Kronecker delta.
With this normalization the differential cross section reads:
\beq
\label{cros}
\frac{d\sigma}{d\Omega}=\frac{1}{64 \pi^2 s} \, \frac{p'}{p} \, |T(ij,ab)|^2~.
\eeq
Finally, a generic meson--baryon S-wave partial wave amplitude is simply given by:
\beq
\label{swave}
T(W)=\frac{1}{8\pi}\sum_{\sigma=1}^2\int d\Omega\, T(W,\Omega;\sigma,\sigma)~,
\eeq
with $\sigma$ the third component of the spin of the baryon in its rest system.

\medskip\noindent {\bf 3.} 
We now present a general scheme that can be applied to any order in the chiral 
calculations. It is simply based on the fact that unitarity, above the pertinent 
thresholds, implies that the inverse of a partial wave amplitude satisfies:
\beq
\label{uni}
\hbox{Im}~T^{-1}(W)_{ij}=-\rho(W)_i \delta_{ij}~,
\eeq
where $\rho_i \equiv q_i/(8\pi W)$ and $q_i$ is the modulus of the c.m. 
three--momentum and the subscripts $i$ and $j$ refer to the physical channels. 
As discussed in the introduction, the $\bar{K}N$ states couple strongly to several 
channels. Also, to be consistent with lowest order CHPT, where all the baryons belonging 
to the same SU(3) multiplet are degenerate, one should consider the whole set of states: 
$K^-p~(1)$, $\bar{K}^0 n~(2)$, $\pi^0 \Sigma^0~(3)$, $\pi^+ \Sigma^-~(4)$, 
$\pi^- \Sigma^+~(5)$, $\pi^0 \Lambda~(6)$, $\eta \Lambda~(7)$, $\eta \Sigma^0~(8)$, 
$K^+\Xi~(9)$, $K^0 \Xi^0~(10)$, where 
between brackets the channel number, to be used in a matrix notation, is given for each 
state. E.g., in this notation, 
$T(W)_{12}$ corresponds to $K^-p \rightarrow \bar{K}^0 n$.
The unitarity relation in eq.(\ref{uni}) gives rise to a cut in the
$T$--matrix of partial wave amplitudes which is usually called the unitarity or right--hand 
cut. Hence we can write down a dispersion relation for $T^{-1}(W)$, in a fairly symbolic 
language: 
\beq
\label{dis}
T^{-1}(W)_{ij}=-\delta_{ij}\left\{\widetilde{a}_i(s_0)+ 
\frac{s-s_0}{\pi}\int_{s_{i}}^\infty ds' 
\frac{\rho(s')_i}{(s'-s)(s'-s_0)}\right\}+{\mathcal{T}}^{-1}(W)_{ij} ~,
\eeq
where $s_i$ is the value of the $s$ variable at the threshold of channel $i$ and 
${\mathcal{T}}^{-1}(W)_{ij}$ indicates other contributions coming from local and 
pole terms as well as crossed channel dynamics but {\it without} 
right--hand cut.\footnote{Albeit we are considering a partial wave amplitude 
$T_{ij}$ as a function of $W$ instead of $s$, in ref.\cite{piN} it was shown that, 
although starting in the  complex--$W$ plane, the final contribution from the 
right--hand cut can be written in the form of eq.(\ref{dis}).} These extra terms
will be  taken directly from CHPT 
after requiring the {\em matching} of our general result to the CHPT expressions. 
Notice also that 
\beq
\label{g}
g(s)_i=\widetilde{a}_i(s_0)+ \frac{s-s_0}{\pi}\int_{s_{i}}^\infty ds' 
\frac{\rho(s')_i}{(s'-s)(s'-s_0)}
\eeq
is the familiar scalar loop integral
\beqa
\label{g2}
g(s)_i&=&\int \frac{d^4 q}{(2\pi)^4}\frac{1}{(q^2-M_i^2+i \epsilon)
((P-q)^2-m_i^2+i\epsilon)}\nonumber \\
&=&\frac{1}{16 \pi^2}\left\{ a_i(\mu)+\log\frac{m_i^2}{\mu^2}+
\frac{M_i^2-m_i^2+s}{2 s}\log\frac{M_i^2}{m_i^2}+\frac{q_i}{\sqrt{s}}
\log\frac{m_i^2+M_i^2-s-2 \sqrt{s}q_i}{m_i^2+M_i^2-s+2\sqrt{s}q_i}
\right\} ,
\eeqa
where $M_i$ and $m_i$ are, respectively, the 
meson and baryon masses in the state $i$. Notice that in order to calculate $g(s)_i$ 
we are using the physical masses both for mesons and baryons since the unitarity 
result in eq.(\ref{uni}) is exact. In the usual chiral power counting,  
$g(s)_i$ is ${\mathcal{O}}(p)$ because the baryon propagator scales as 
${\cal O}(p^{-1})$. However, in its full relativistic form, as the one
shown in eq.(\ref{g2}), there is also a term $a_i(\mu)+\log
m_i^2/\mu^2$ which is not homogeneous 
in the external four--momenta of the mesons. This kind of problems when calculating loops 
relativistically in CHPT in the baryon sector are known for long \cite{gl}. Nevertheless, 
we will treat in the following the $g(s)_i$ functions as ${\mathcal{O}}(p)$ since 
the first two constant terms in eq.(\ref{g2}) give rise only to local vertices that 
can be reabsorbed order by order in the CHPT counterterms
(which is similar to the procedure advocated in \cite{bech}). 
Let us note that the important 
point here is to proceed systematically guaranteeing that $\tc$ is
free of the right--hand cut  and matching simultaneously with the CHPT
expressions. Consistency of course 
requires to proceed in the same way whereever this formalism is applied. 

\medskip\noindent
We can further simplify the notation by employing a matrix formalism. We  
introduce the 
matrices $g(s)={\rm diag}~(g(s)_i)$, $T$ and ${\mathcal{T}}$, the latter defined in 
terms 
of the matrix elements $T_{ij}$ and ${\mathcal{T}}_{ij}$. In this way, 
from eq.(\ref{dis}), the $T$-matrix can be written as:
\beq
\label{t}
T(W)=\left[I+{\mathcal{T}}(W)\cdot g(s) \right]^{-1}\cdot {\mathcal{T}}(W)~.
\eeq

\medskip\noindent
In this letter we are considering the lowest order (tree level) CHPT amplitudes as 
input. Hence,  expanding the previous equation, we will have:
\beq
\label{ext}
T(W)=\tc(W)-\tc(W) \cdot g(s) \cdot \tc(W) + \ldots ~,
\eeq 
where the ellipsis denotes terms of higher order.
At lowest order $T_1(W)\equiv T(W)+{\mathcal{O}}(p^2)$ does not 
contain loops and thus we will have up-to-and-including ${\mathcal{O}}(p)$:
\beq
\label{mat}
\tc_1(W)=T_1(W)~,
\eeq
with the subscript indicating the chiral order. Thus our final expression for the 
$T$-matrix, taking as input the lowest order CHPT results, has the form
\beq
\label{fin}
T(W)=\left[ I + T_1(W) \cdot g(s) \right]^{-1} \cdot T_1(W) ~.
\eeq
It is straightforward how this procedure can be applied to any order 
of a given CHPT calculation in order to provide this perturbative result with the resummation of 
the right--hand cut contributions. For instance, if we were given the next-to-leading order 
CHPT amplitudes then we would have:
\beq
\label{ex2}
T_1+T_2=\tc_1+\tc_2~,
\eeq   
with $\tc_1(W)\equiv T_1(W)$ already given. As before, the subscripts refer to the chiral 
order. In the same way, up to  ${\mathcal{O}}(p^3)$:
\beq
\label{ex3}
T_1+T_2+T_3=\tc_1+\tc_2+\tc_3-\tc_1 \cdot g \cdot \tc_1 ~,
\eeq
and so on. In this way the matrix $\tc$ is given order by order in the chiral counting 
and then it is substituted in eq.(\ref{t}) to finally obtain the matrix of 
partial wave amplitudes, $T$.
Notice as well that  this scheme is general enough to be 
applied in other processes where the right--hand cut contributions are far from being 
perturbative, such as nucleon--nucleon scattering.

\medskip\noindent
It is important to stress  that if instead of considering pure CHPT amplitudes 
one also includes resonances (as in refs.\cite{piN,bku,jamin}),  the
chiral counting employed so far has to be changed to 
match our general expression eq.(\ref{t}) to the input. This is so because the inclusion 
 of explicit resonance fields already at the tree--level implies an infinite
tower of chiral orders while the loops are still calculated perturbatively. For 
instance, if as in ref.\cite{piN} 
our input $T_R(W)$ consists of lowest order CHPT plus resonances plus
{\it one-loop} contributions 
calculated at ${\mathcal{O}}(p^3)$ in CHPT, we will have from eq.(\ref{ext}):
\beq
\label{res}
T_R=\tc-\tc_1 \cdot g(s) \cdot \tc_1 ~,
\eeq
from which one isolates $\tc(W)$ to be substituted in eq.(\ref{t}). 

\medskip\noindent
Next, we discuss the subtraction constants  $a_i(\mu)$ appearing in eq.(\ref{g2}).
In order to have some estimate of the value of the $a_i(\mu)$, we consider the comparison 
between the one--loop function $g(s)_i$ and an approximation to it given by the calculation of 
the loop integral in eq.(\ref{g2}) by means of a cut--off,  as done in ref.\cite{oset}. 
The calculation of $g(s)_i$ given in eqs.(\ref{g},\ref{g2}) by means of a finite cut-off is 
only approximative because the real part of $g(s)_i$ derived in 
this way does not fulfill the dispersion relation given in eq.(\ref{g}). A 
straightforward comparison between both regularization schemes gives: 
\beq
\label{com}
a_i(\mu)=-2\log \left(1+\sqrt{1+\frac{m_i^2}{\mu^2}} \right)+...~,
\eeq
where the ellipses indicate higher order terms in the non--relativistic expansion and 
also powers of $M_i/m_i$. For instance, for an average mass of the octet of 
$J^{P}=\frac{1}{2}^+$ baryons, $m_i = 1.15$ GeV and the value of $\mu=630$ MeV considered 
in ref.\cite{oset}, one has $a\simeq -2.3$. A ``natural'' value for
$a$ is around $-2$ because one expects $\mu$ to be somewhere in the region of
the first (meson) resonance, which is the $\rho (770)$. However, one should remember 
that this is not the 
most general case. In fact, a priori, it could be that the value of the subtraction 
constant $a(\mu)$ were such that an unphysical value of the cut--off should be necessary 
to reproduce its value. For instance, it is obvious from the previous equation that 
$\mu=\infty$ for $a(\mu)=-2 \log 2\simeq -1.39$. In fact, there is a strong sensitivity 
of the value of the cut--off $\mu$ in terms of $a(\mu)$ due to exponential dependence,
cf. eq.(\ref{com}).  All this clearly shows that the 
appropriate way to proceed is the one presented here, namely to include an explicit 
subtraction constant. Due to the renormalization group invariance any change in the 
scale $\mu$ is reabsorbed by a change in $a(\mu)$ such that $a(\mu')-a(\mu)=
\log\mu'^2/\mu^2$, cf. eq.(\ref{g2}), and the amplitudes $T(W)$ are, of course, 
scale--independent.  With that in mind,  we will fix the  scale to $\mu=630$ MeV in what 
follows.

\medskip
 
\noindent {\bf 4.} Our calculated amplitudes $T(W)$ depend on the values of the 
parameters $m_0$ and $F_0$ (coming from the lowest order CHPT amplitudes, eqs.(\ref{amp})) 
and on the $a_i(\mu)$, the subtraction constants appearing in the 
functions $g(s)_i$, cf. eq.(\ref{g2}). Since $m_0$ represents the average mass of 
the considered baryon octet in the chiral limit, to lowest order
$m_0\approx 1.15$ GeV, which is nothing but the average of the physical masses of the baryons
 belonging to this multiplet. With respect to the value of $F_0$, the situation is 
still rather controversial. Notice that since we are considering SU(3) CHPT 
the $F_0$ appearing in eqs.(\ref{amp}) must be smaller than the
two--flavor decay constant $F_0\equiv F_0^{(2)}$ \cite{stern}. 
{}From ref.\cite{gl2} one has $F_\pi/F_0^{(2)}=1.057\pm 0.012$, 
with $F_\pi=92.4$ MeV the weak pion decay constant, that is $F_0^{(2)}\simeq 87.4\,$MeV.
However, the precise value of $F_0$ in the three flavor case strongly depends on the 
value of the low--energy constant $L_4$ which is not well known \cite{stern}, see also 
ref.\cite{om} for a detailed discussion about this issue comparing different works. 
In ref.\cite{gl2} one can find the ratio $F_\pi/F_0=1.07\pm 0.12$ which results after 
estimating the order of magnitude of $L_4$. From the value of $L_4$ given in  
ref.\cite{om} one obtains that $F_0= 72.2$ MeV, which is on the lower side of  
the results of ref.\cite{gl2}. To summarize this topic, one expects the value of
$F_0$ in the range from $70$ to $90\,$MeV.
With respect to the subtraction constants $a_i(\mu)$, 
we will use one single (average) value $a(\mu)$ for all of them.
This is mostly done to reduce the number of free parameters of the approach. In fact,
we have checked that considering also the general case with different values 
for the subtraction constants $a_i(\mu)$, neither the quality of the reproduction of 
the data nor the conclusions of our study  change in a significant way.

\medskip
\noindent
We first consider the set of {\em natural} 
values $m_0=1.15$ GeV, $F_0=86.4$ MeV and $a(\mu)=-2$ as discussed above, which 
we will call set~II in the following. In fig.2, the results of our approach for this 
case for several $K^-p$ cross sections from threshold up to 250 MeV of the 
incoming kaon three--momentum in the laboratory frame, $k_L$, are shown by the dashed lines. 
Thus, a quite fair reproduction of the scattering date is obtained in a very natural way. 
Together with this scattering data we also consider, as in ref.\cite{siegel}, the 
three well measured threshold ratios \cite{nowak,tovee}  of the $K^-p$ system:
\beqa
\label{rat}
\gamma&=&\frac{\Gamma(K^-p \rightarrow \pi^+\Sigma^-)}{\Gamma(K^-p \rightarrow 
\pi^- \Sigma^+)}=2.36\pm 0.04~, \nonumber \\
R_c&=&\frac{\Gamma(K^-p\rightarrow \hbox{charged particles})}{\Gamma(K^-p 
\rightarrow \hbox{all})}=0.664\pm 0.011\nonumber~, \\
R_n&=&\frac{\Gamma(K^-p \rightarrow \pi^0 \Lambda)}{\Gamma(K^-p \rightarrow 
\hbox{all neutral states})}=0.189 \pm 0.015~.
\eeqa
These threshold ratios imply additional tight constraints on the $\bar{K}N$ 
interactions, as first stressed in ref.\cite{siegel}. For the set~II, the
following values of the ratios are obtained: $\gamma=2.05$, $R_c=0.624$ and $R_n=0.264$, 
quite close to the experimental values. However, one has to say that the value of the 
ratio $\gamma$ is very sensitive to small variations of the subtraction constant 
$a(\mu)$. For instance a relative change of less than a $10\%$ in $a(\mu)$ gives rise to 
a change of more than a $50\%$ in $\gamma$. This can be traced back to the movement in 
the complex plane of the pole position of the $\Lambda(1405)$ as seen in more detail 
below. Such a strong sensitivity has also been observed in cut--off
schemes, see e.g. \cite{kaiser,oset}.
\begin{figure}[t]
\centerline{
\epsfysize=12cm
\epsffile{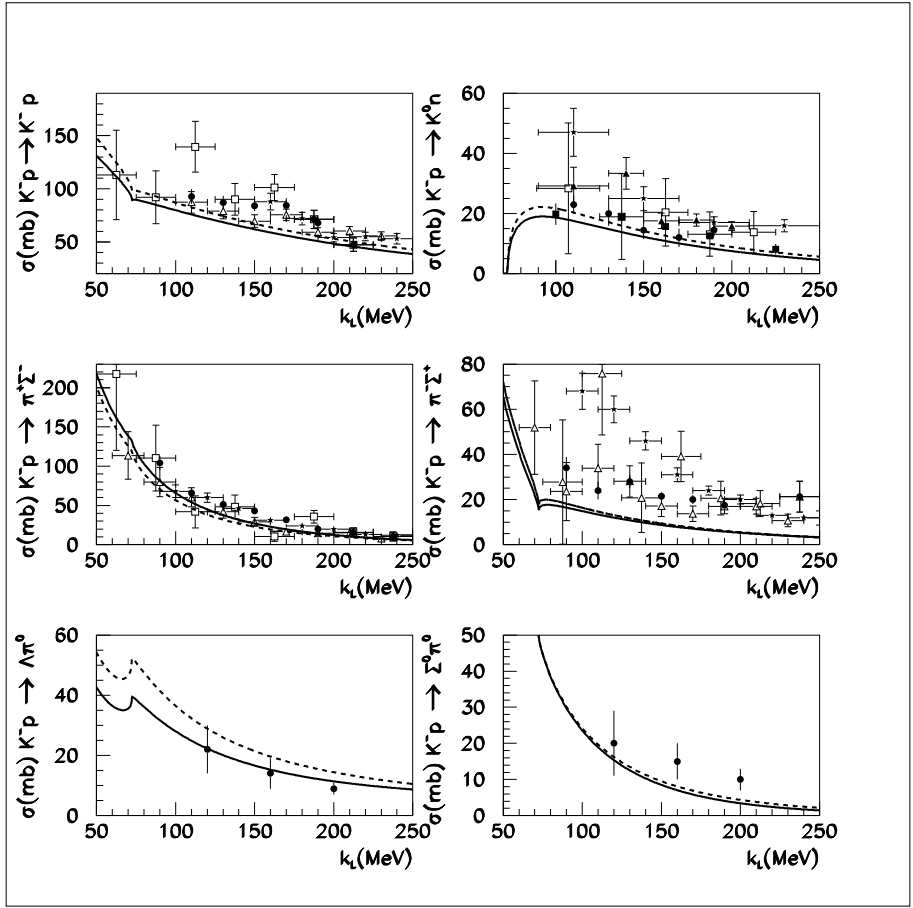}
}
\vspace{-0.1cm}
\caption{Scattering data in the low energy region. The solid and dashed  lines refer
to the parameter sets~I and II, respectively, as explained in the text. Experimental 
data: empty squares \cite{hump}, full circles \cite{kim}, empty triangles \cite{sakitt}, 
full squares \cite{kittel}, full triangles \cite{evans} and stars \cite{cibo}.}
\vspace{-0.3cm}
\end{figure}

\medskip\noindent
We now consider the case in which all the parameters $m_0$, $F_0$ and $a(\mu)$ 
are let free and are fitted to the scattering data below $k_L=150$ MeV and to the threshold
ratios. These parameters are called set~I for short.
For energies higher than 150 MeV there are indications that 
the P-waves become important \cite{lutz,cibo}.\footnote{We note that our approach 
as formulated here can  directly be applied  to higher partial waves.} 
The resulting values for the parameters 
are: $m_0=1286$ MeV, $F_0=74.1$ MeV and $a(\mu)=-2.23$, leading to the
threshold ratios:
\beq
\gamma=2.33~, \quad R_c=0.645~, \quad R_n=0.227~. 
\eeq
It is remarkable that the 
values of the parameters $m_0$, $F_0$ and $a(\mu)$ turn out to be in the range of the 
expected values discussed above from other sources of phenomenology. On the other hand, 
although the reproduction of the scattering data (as given by the solid lines in fig.2)
is very similar in quality to that 
obtained with the previous fixed values for $m_0$, $F_0$ and $a(\mu)$ the ratios are 
better described. However, the fitted values for the parameters should
only be considered indicative since higher order contributions are expected to give 
non--negligible contributions. We have checked that by allowing
different values for the decay  
constant $F_0$ for each channel as an indication of higher order contributions. In fact, 
there is an improvement in the quality of the fit and in particular 
the cross section $\sigma(K^-p\rightarrow \pi^- \Sigma^+)$ is then better 
reproduced with a resulting  $\chi^2$ per degree of freedom lower than 2. 
With respect to this $\chi^2$, it is worth to note 
that the different sets of data are not always mutually consistent,
as can be seen in fig. 2.  However, we consider  the achieved global description of the 
scattering data and threshold ratios as quite satisfactory  
(later on we will also show a nice reproduction of the 
$\pi^-\Sigma^+$ event distributions around the position of the
$\Lambda(1405)$ resonance) in terms of the global parameters $m_0$, $F_0$ and $a(\mu)$ 
both for the set of fixed {\it{natural}} values, set~II, or when taking them as 
free parameters, set~I.

\medskip\noindent 
Next, we  consider the $K^- p$ scattering length. For set~I, we obtain 
$a_0=-0.58+ i 1.19$ fm (with all particle masses at their physical
values)  and 
$a_0=-0.53+ i 0.95$ fm, the latter in the isospin limit (taking equal  
masses of all the particles in an isospin multiplet when calculating the 
$g(s)_i$ functions). These values are rather similar 
to those obtained from set~II: $a_0=-0.75 + i1.21$ fm and $a_0=-0.62 + i 0.96$ fm, 
respectively. For comparison, we give  the recent results for $K^- p$ scattering lengths 
determined from kaonic hydrogen $X$--rays \cite{iwasaki}, $a_0=(-0.78\pm 0.15 \pm 0.03) 
+ i \,(0.49\pm 0.25 \pm 0.12)$ fm and the value of \cite{admartin}, 
$a_0=(-0.67 \pm 0.10)+i\, (0.64\pm0.10)$, calculated from the isospin scattering lengths 
of ref.\cite{brmartin}. The agreement is quite good. The rather stable  value 
of the calculated imaginary part of the scattering length of about 1~fm is 
consistent with the findings of refs.\cite{kaiser,oset}.

\medskip\noindent
Finally, some remarks concerning the importance of the various
contributions of our input $T_1(W)$ in eq.(\ref{fin}) are necessary.
We have checked that the direct and crossed terms, see fig.1, 
although negligible at energies around 1.3 GeV (as it is expected since they are 
subleading in the 
non--relativistic counting) rapidly increase in 
magnitude with energy and for some channels they can be as large as  $\sim 20\%$ 
of the dominant seagull term around 1.5 GeV. Note that in
ref.\cite{oset} only the seagull term, further restricted to its 
non--relativistic limit, was considered. In ref.\cite{kaiser} some 
relativistic corrections were included since they considered next--to--leading order 
vertices in the heavy baryon CHPT formalism. However, we have to stress that the 
relativistic expression for the $g(s)_i$ functions eq.(\ref{g2}) is of outmost 
importance in order to reproduce the $\bar{K}N$ dynamics. Indeed, when considering 
only its leading term in a combined non--relativistic 
$(|\vec{p}\,|/m_i)$ and $M_i/m_i$  expansion, the results change dramatically. This is due 
to the large mass of the kaons as 
compared to the mass of the baryons which makes the functions $g(s)_1$ and $g(s)_2$
differ sizeably from their leading non--relativistic expressions, particularly  below 
and close to the $\bar{K}N$ threshold. Let us note that the appearance of the 
$\Lambda(1405)$, see below, will result 
from a cancellation of the denominator of the matrix elements of eq.(\ref{fin}). 
This cancellation is a clear interplay between 
the tree level vertices and the $g(s)_i$ functions and hence such drastic changes 
in $g(s)_i$, when considering the leading non--relativistic limit, make the final 
results change drastically, too.

\medskip\noindent 
{\bf 5.} We now address the problem of calculating the $\pi^-\Sigma^+$ event 
distribution 
in the vicinity of the $\Lambda(1405)$. Typically this observable has been 
calculated assuming that the process is dominated by the $\pi\Sigma$ $I=0$ system, so 
that it is proportional to the strong $I=0$ $T_{\pi\Sigma \rightarrow \pi\Sigma}$ S-wave 
amplitude. However, we want to stress here that this is an oversimplification 
since the threshold of the $\bar{K}N$ system is very close to 1.4 GeV and 
one should consider from the beginning a coupled channel scheme. We do this here 
following the approach given in ref.\cite{om} for the study of the $J/\Psi$ decays to 
a vector ($\phi(1020)$ or $\omega(782)$) and to two pions or kaons. That problem 
is very similar to the one here since the two pions or kaons couple strongly around 
the region of the $f_0(980)$ (as here the $\pi\Sigma$ and the $\bar{K}N$ states around the 
mass of the $\Lambda(1405)$). Thus, considering that the $\pi\Sigma$ state originates 
form a generic S-wave $I=0$ source (for instance, a $\Lambda(1405)$ resonance) one has 
to introduce a 10--component vector of 
{\em scalar form factors} (for details, see ref.\cite{om}):
\beqa
\label{diss}
F&=&[I+\tc \cdot g]^{-1}\cdot R~, \nonumber \\
R^T&=&\left(\begin{array}{cccccccccc}
r_1, & r_1, & r_2, &  r_2,  & r_2, & 0, &  0, &  0, &  0, & 0
\end{array}
\right)~,
\eeqa
where we have set those elements of $R$ equal zero that correspond to the states different to 
$\bar{K}N$ and $\pi\Sigma$ which are assumed to be the dominant ones in this energy 
region. Notice that in writing $R$ we have taken advantage of the fact that the
source has isospin zero ($I=0$), 
therefore only two different constants appear in this limit, denoted $r_1$ for the $K^-p$ 
and $\bar{K}^0n$ channels and  $r_2$ for $\pi^0 \Sigma^0$, $\pi^+\Sigma^-$ and 
$\pi^- \Sigma^+$. From the previous equation the $\pi^- \Sigma^+$ event distribution, which 
is not normalized, is then given by:
\beq
\label{disfor}
\frac{d N_{\pi^- \Sigma^+}}{d E}=|F(5)|^2 \,  p_{\pi^-\Sigma^+}~.
\eeq
The usual approach of 
considering only the $T_{\pi\Sigma \rightarrow \pi\Sigma}$ $I=0$ S-wave amplitude 
results from eq.(\ref{diss}) when taking $r_1=0$ (compare with eq.(\ref{t})), as 
we have also checked numerically. Notice that in a sufficiently narrow energy 
region around 1.4 GeV the lowest order 
CHPT results can be well approximated by constants, in the same way as we are also doing 
in eq.(\ref{diss}) with respect the vector $R$.
In fig. 3. we compare our calculated event distribution (shaded area) for set~I with 
the experimental data from ref.\cite{hem}. The error given to the experimental points 
is $\sqrt{N}\times 1.6$ when considering also the uncertainties in the event selection 
\cite{dalitz}. The agreement is rather satisfactory. Notice that this event 
distribution is a result of the already fixed strong scattering amplitudes.
\begin{figure}[htb]
\centerline{
\epsfysize=7cm
\epsffile{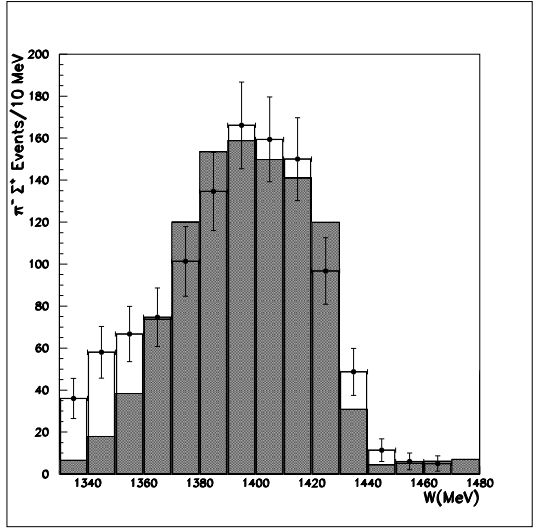}
}
\vspace{-0.1cm}
\caption{Event distribution in the region of the $\Lambda (1405)$ for set~I
(shaded area) in comparison to the data from \protect\cite{hem}.}
\vspace{-0.3cm}
\end{figure}

\medskip\noindent
We finally discuss the poles in the complex energy plane for 
the different Riemann sheets. In discussing the $\Lambda(1405)$ resonance the closest 
thresholds are 
the ones corresponding to the $\pi\Sigma$ and $\bar{K}N$ states. Hence, one has to 
consider generally four sheets. The physical one labelled by $(+,+)$, the second 
$(-,+)$, the third $(-,-)$ and the fourth sheet $(+,-)$. The signs inside the 
brackets refer to the sign ambiguity 
present in the definition of the modulus of the c.m. three--momentum for the 
channels $\pi\Sigma$ and $\bar{K}N$, respectively. In this way, the `$+$' sign 
indicates that the imaginary part of the modulus of the three--momentum is always 
positive in the whole $s$--plane and the `$-$' sign refers to the contrary (unphysical 
sheet). When looking at the pole structure in the unphysical Riemann sheets one 
realizes  the presence of three poles corresponding to the three expected neutral 
 members of a $J^P=\frac{1}{2}^-$ baryon nonet (in the isospin limit, two of them with 
$I=0$ and another one with $I=1$). We also find that the pole positions change 
appreciably from one sheet to the other, which is a clear indication of a large 
meson--baryon component. For the second and third sheets, 
which are the closest ones to the physical sheet, we have the following pole positions. 
Sheet~II: $(1379.2-i\,27.6)\,$MeV, $(1433.7-i\,11.0)\,$MeV ($I=0$) and 
$(1444.0-i\,69.4)\,$MeV ($I=1$). Sheet~III: $(1346.2-i\,3.0)\,$MeV 
and $(1411.9-i\,48.8)\,$MeV ($I=0)$ and $(1419.0-i\,41.7)\,$MeV ($I=1$). Note that 
the $\Lambda(1405)$ resonance is described by two poles on  sheets II and III with 
rather different imaginary parts indicating a clear departure from the Breit-Wigner situation 
for this resonance as already stated long time ago, see e.g. 
refs. \cite{hem,dalitz}. It is also worth to remark 
the presence of another pole with $I=0$ very close to the threshold of 
the $\pi\Sigma$ states with 
a mass inside the range of energies covered by fig.3. In addition, 
since we have seen that the resulting value of the subtraction constant can easily be 
explained as coming from a cut--off with a 
{\it natural} value, one should conclude 
that, at least, this set of poles largely corresponds to meson--baryon resonances. 
However, one should  include explicit resonance fields (which can be straightforwardly 
done in this approach) in order to finally assess the absence or the presence of 
small preexisting baryon resonance components in their wave 
functions, analogously as it has been done in the meson--meson sector 
in refs.\cite{oo,jamin}. Finally, the $S=0$ \cite{kaiser2,oset2} and $S=-2$ S-wave meson--baryon 
amplitudes should be studied analogously in order to complete the assignment of resonances 
belonging to the aforementioned $\frac{1}{2}^-$ baryon nonet.    

\medskip\noindent
{\bf 6.} We have derived a general and systematic method to resum the right--hand cut 
contributions from a perturbative (chiral) series. This method should be applied 
whenever 
these contributions become far from being perturbative both in the low (or higher) 
energy regimes, as for instance in the S--wave kaon--nucleon interactions for 
strangeness $S=-1$ as discussed in detail here. The theory can be seen as a 
reformulation in more general terms of the approach already applied to study the 
meson--meson scattering in refs.\cite{jamin,oo,ww}. Dispersion relations are used to 
perform
the necessary resummation of the chiral perturbation theory amplitudes
given at any order. These CHPT amplitudes are then incorporated in our 
general expressions for the partial wave amplitudes by requiring the 
matching  between both results in a well defined chiral power counting. 
Here, we have simply considered the lowest order
(tree level) CHPT approximation as our starting point. A good description of the 
scattering data  in the $K^- p$, $\pi \Sigma$ and $\pi \Lambda$ channels
as well as for the threshold branching ratios is obtained. In
addition, we have given an improved theoretical prescription to
calculate the $\Sigma \pi$ event distributions in the region of the
$\Lambda (1405)$ leading to a good reproduction of the data. We have also
discussed the pole structure in that energy region. In the future, one
should extend these considerations in three ways. First, a second or
third order (relativistic) CHPT amplitude should be used as
input and higher partial waves should be included. 
Second, one should consider other strangeness channels 
(in particular $\pi N$ scattering has to be treated systematically, see \cite{piN})
and also include explicit resonance fields (as it is possible in this
framework). This
is of particular interest for the question whether the $\Lambda
(1405)$ is a pure meson--baryon boundstate or has a small
``preexisting'' component. Third, 
the extension to electromagnetic kaon production off nucleons is of interest 
due to the data from ELSA and TJNAF.

\medskip

\noindent {\bf Acknowledgements}

\medskip

\noindent
The work of J.A.O. was supported in part by funds from
DGICYT under contract PB96-0753 and from the EU TMR network Eurodaphne, contract no. 
ERBFMRX-CT98-0169.

\bigskip

\end{document}